\documentclass[preprint]{elsarticle}

\usepackage{graphicx}

\begin{document}

\title{
 Proposal of a Checking Parameter in the Simulated Annealing Method Applied to the Spin Glass Model
}

\author{
Chiaki Yamaguchi%
}

\ead{chiaki77@hotmail.com}

\address{
Kosugichou 1-359, Kawasaki 211-0063, Japan
}

\begin{abstract}
 We propose a checking parameter utilizing the breaking of the Jarzynski equality
 in the simulated annealing method using the Monte Carlo method.
 This parameter is based on the Jarzynski equality.
 By using this parameter, to detect that
 the system is in global minima of the free energy
 under gradual temperature reduction is possible.
 Thus, by using this parameter, one is able
 to investigate the efficiency of annealing schedules.
 We apply this parameter to the $\pm J$ Ising spin glass model.
 The application to the Gaussian Ising spin glass model is also mentioned.
 We discuss that the breaking of the Jarzynski equality is induced
 by the system being trapped in local minima of the free energy.
 By performing Monte Carlo simulations of the $\pm J$
 Ising spin glass model and a glassy spin model proposed by Newman and Moore, 
 we show the efficiency of the use of this parameter.
\end{abstract}


\maketitle

\section{Introduction}

The studies of spin glass models have been 
 widely done\cite{KR, MPV}.
 The spin glass models have the randomness and the frustration.
 The combination of the randomness
 and the frustration causes various
 interesting dynamics as well as the static properties.
 For the spin glass model, there is a problem that
 it is difficult for the system to reach global minima of the free energy
 by using the Monte Carlo method,
 although the Monte Carlo method is known as a powerful method
 for investigating spin models.
 When the system has reached the global minima of the free energy
 by the Monte Carlo method,
 the system is in equilibrium or the ground state.
 On the other hand, when the system has not
 reached the global minima of the free energy by the Monte Carlo method,
 the system is in non-equilibrium.

We propose a checking parameter in the simulated annealing method
 using the Monte Carlo method.
 The simulated annealing method originally uses the Metropolis method\cite{MRRTT, KGV}
 and is originally made for the ground-state search\cite{KGV}.
 The simulated annealing method\cite{KGV, GG}
 performs gradual temperature reduction (annealing).
 It is known that,
 if $T(k) \ge \frac{c}{\log (1 + k)}$ is applied as the time schedule (annealing schedule),
 the system always goes in the ground state for any model,
 where $T$ is the temperature, $c$ is a constant, and $k$ is the number of
 site replacements\cite{GG}.
 However, this time schedule is not practical for simulation time.
 In addition, because the difficulty for finding the ground state
 depends on each model, each proper annealing schedule can also depend on each model.
 In the cases of the complex systems such as the spin glass model,
 if annealing schedules are not appropriate,
 the systems go in local minima of the free energy,
 and one can not obtain the physical quantities in equilibrium or the ground state.
 If appropriate annealing schedules are chosen,
 the systems go in global minima of the free energy,
 and one can obtain the physical quantities in equilibrium or the ground state.
 Therefore, when performing the simulated annealing method,
 one has to choose appropriate annealing schedules.

 The understandings for solving optimization problems
 by using annealing processes are recently becoming significant
 in relation to the D-Wave chip\cite{Jetal}.
 This chip is designed to perform a quantum annealing
 for solving optimization problems.
 The present study is related to a thermal annealing, but
 the relationship with a quantum annealing\cite{FGGS} is straightforward.
 The application to the quantum annealing is mentioned in \S\ref{sec:5}.

 There are two interests for investigating the spin glass model
 by using the Monte Carlo method at least.
 One is to clear the nature of the material, and another is
 to find a powerful Monte Carlo method.
 The spin glass model also works as a test bed for optimization methods\cite{KGV}.
 This study has a relationship with both of the two interests, because
 the study for gradual temperature reduction is
 proposed as an optimization method called the simulated annealing method\cite{KGV, GG}
 and  the study for gradual temperature reduction is also done for
 investigating dynamical features of the spin glass model\cite{D}.

 We apply the Jarzynski equality to the spin glass model.
 The Jarzynski equality is an equality that connects the work
 in non-equilibrium and the ratio of the partition functions\cite{J1, J2}.
 The work is performed in switching an external parameter of the system.
 The Jarzynski equality for temperature-change process
 is also proposed\cite{C1, OKN}.
 We use the Jarzynski equality for temperature-change process.
 The Jarzynski equality is also derived in the Markov process with discrete time
 in Ref.\cite{C2}, and it is pointed out in Ref.\cite{C2} that
 the Metropolis method\cite{MRRTT}
 based on the Markov process with discrete time
 is a suited example for applying the Jarzynski equality.

 We propose a checking parameter in this article.
 By using this checking parameter,
 to detect that the system is in global minima of the free energy is possible.
 The checking parameter means that, by using this parameter,
 one can check whether annealing schedules are appropriate or not.
 The meaning of the value of the present checking parameter
 is different from that of the study of the energy decrease as the temperature decreases,
 although this parameter uses the values of the energies.
 In the case of the study of the energy decrease,
 one can see that the energy does not decrease for long Monte Carlo time, however,
 one can not determine whether
 the system is in global minima of the free energy or not.
 On the other hand,
 because the value of a quantity estimated by the Monte Carlo method
 and the exact analytical value of the quantity are compared,
 by using the present checking parameter,
 one can determine whether
 the system is in global minima of the free energy or not.

 In this article, in order to show the features of the present checking parameter,
 we apply the Glauber dynamics\cite{D, Og},
 although other Monte Carlo methods,
 which include the Metropolis method\cite{MRRTT},
 based on the Markov process with discrete time are also applicable.
 The Glauber dynamics is suited to the study
 of the dynamical features of physical systems\cite{D, Og}.
 If one uses the present checking parameter as a powerful optimization method,
 applying more powerful Monte Carlo methods\cite{HN, N},
 instead of the Glauber dynamics or the Metropolis method, would be necessary.

 We apply the present parameter to the $\pm J$ Ising spin glass model\cite{KR}.
 The present technique is also applicable to
 the Gaussian Ising spin glass model\cite{MPV}.
 The application to the Gaussian model is also mentioned in this article.

There are previous studies for investigating
 the relationship between the spin glass model and the Jarzynski equality
 in Refs.\cite{OKN, ON, Y, O}.
 We describe the difference between this study and the previous studies.
 We propose a checking parameter.
 By using this parameter, one is able
 to investigate the efficiency of annealing schedules.
 In the previous studies, this parameter is not mentioned,
 the breaking of the Jarzynski equality is not mentioned,
 and the treatment under a uniform external magnetic field is also not mentioned.
 In this article, these are described.

 In addition, in order to confirm the behavior of the present checking parameter,
 a glassy spin model \cite{NM} by Newman and Moore is also investigated.

This article is organized as follows.
 The present checking parameter is explained in \S\ref{sec:2},
 and the breaking of the Jarzynski equality is discussed in \S\ref{sec:3}.
 The results of Monte Carlo simulations are given in \S\ref{sec:4}.
 The concluding remarks of this article are described in \S\ref{sec:5}.

\section{A Checking Parameter} \label{sec:2}

We investigate the $\pm J$ Ising spin glass model.
The Hamiltonian ${\cal H}$ for Ising spin glass models
 is given by  \cite{KR, MPV}
\begin{equation}
 {\cal H} = - \sum_{\langle i, j \rangle} J_{i, j} S_i S_j - h \sum_i S_i \, ,
\end{equation}
 where $\langle i, j \rangle$ denotes nearest-neighbor pairs, $S_i$ is
 a state of the spin at site $i$, $S_i = \pm 1$, and
 $h$ is an external magnetic field.
 The value of $J_{i, j}$ is given with a distribution $P ( J_{i, j})$.
 The distribution $P^{(\pm J)} ( J_{i, j} )$ of $J_{i, j}$ for the $\pm J$ model
 is given by \cite{KR}
\begin{equation}
 P^{(\pm J)} ( J_{i, j})
 = \frac{1}{2} \, \delta_{ J_{i, j}, J} + \frac{1}{2} \, \delta_{ J_{i, j}, - J} \, ,
 \label{eq:PpmJJij}
\end{equation}
 where $\delta$ is the Kronecker delta, $J > 0$, and
 $J$ is the strength of the exchange interaction between spins.

We explain the Jarzynski equality\cite{J1, J2, C2}.
 We consider a non-equilibrium process of $\lambda_t$ from $\lambda_0$ to $\lambda_\tau$.
 $\lambda_t$ is an externally controlled parameter, and $t = 0, 1, 2, \ldots, \tau$.
 The initial and final states in equilibrium are assumed,
 and the states in the process from $\lambda_0$ to $\lambda_\tau$ are in non-equilibrium.
 The Jarzynski equality is equivalently given by \cite{J1, J2, C2} 
\begin{equation}
 \overline{e^{- \beta W}} = \frac{Z (\beta, \lambda_\tau )}{Z (\beta, \lambda_0 ) } \, ,
 \label{eq:Jeft0}
\end{equation}
 where 
 $W$ is the work performed in the process from $\lambda_0$ to $\lambda_\tau$, 
 $\beta$ is the inverse temperature of the reservoir,
 $\beta = 1 / k_B T$, $T$ is the temperature,
 and $k_B$ is the Boltzmann constant.
 The overbar indicates an ensemble average over all possible paths through phase space. 
 $Z$ is the partition function given by $Z = \sum \exp (- \beta {\cal H})$.
 The left-hand side of Eq.~(\ref{eq:Jeft0}) is the 
 non-equilibrium measurements, and
 the right-hand side of Eq.~(\ref{eq:Jeft0}) is the 
 equilibrium information.
 By using Eq.~(\ref{eq:Jeft0}),
 to extract the equilibrium information from
 the ensemble of non-equilibrium is possible.
 The work $W$ is given by \cite{C2}
\begin{equation}
 W = \sum^{\tau - 1}_{t = 0} [ E ( i_t, \lambda_{t + 1} ) - E (i_t, \lambda_t ) ] \, , \label{eq:W}
\end{equation} 
 where $E (i_t, \lambda_t )$ is the energy at state $i_t$
 under  the externally controlled parameter $\lambda_t$.
 The Jarzynski equality for temperature-change process
 is proposed in Refs.\cite{C1, OKN}.
 In Ref.\cite{C2}, the Jarzynski equality is derived
 in the Markov process with discrete time.
 Then, the inverse temperature $\beta$
 and the energy $E (i_t, \lambda_t )$ always appear as a couple.
 Therefore, the Jarzynski equality also holds for process
 of $\beta_t E (i_t )$ instead of $\beta E (i_t, \lambda_{t})$, where
 $\beta_t$ is the inverse temperature with discrete time $t$,
 and $E (i_t )$ is the energy at state $i_t$ with discrete time $t$.
 Then, the Jarzynski equality for the process of the inverse temperature $\beta_t$
 from $\beta_0$ to $\beta_\tau$ is given by
\begin{equation}
 \overline{e^{- \Upsilon }} =  \frac{Z (\beta_\tau )}{Z (\beta_0 ) } \, , \label{eq:Jeft1}
\end{equation}
 where $\Upsilon$ is a pseudo work given by
\begin{equation}
 \Upsilon = \sum^{\tau - 1}_{t = 0} ( \beta_{t + 1} - \beta_t ) E ( i_t ) \, . \label{eq:mu}
\end{equation}
 Only the values of the energies just before the temperature changes
 contribute to the value of $\Upsilon$.

We consider a quantity $[ \overline{e^{- \Upsilon }} ]_R$, where $[ \, ]_R$
 is the random configuration average for exchange interactions.
 $[ \overline{e^{- \Upsilon }} ]_R$ is a quantity for $\Upsilon$, and
 $[ \overline{e^{- \Upsilon }} ]_R$ is also a quantity for the ratio of the partition functions.
 Note that, strictly speaking, 
 the quantity $[ \overline{e^{- \Upsilon }} ]_R$ is not the free energy difference,
 because the free energy difference is given by 
 $- \frac{1}{\beta_\tau} [ \ln Z (\beta_\tau ) ]_R + \frac{1}{\beta_0} [ \ln Z (\beta_0 ) ]_R$
 and is not a function for $[ \overline{e^{- \Upsilon }} ]_R$, but 
 the quantity $[ \overline{e^{- \Upsilon }} ]_R$ is useful as described in this article.
 When $\beta_0 = 0$ and $\beta_\tau = \beta$, 
 by applying Eq.~(\ref{eq:Jeft1}) to the $\pm J$ Ising spin glass model, we obtain 
\begin{eqnarray}
 [ \overline{e^{- \Upsilon }} ]^{(\pm J)}_R &=&
  \frac{1}{2^{N_B}} \sum_{\{ J_{i, j} \} } \frac{\sum_{\{ S_i \} }
 e^{\beta \sum_{\langle i , j \rangle } J_{i, j } S_i S_j  + \beta h \sum_i S_i  } }{2^N}
 \nonumber \\ 
  &=& \exp \{ N_B \ln [ \cosh (\beta J) ]  + N \ln [ \cosh (\beta h) ] \} \, , \label{eq:mu2}
\end{eqnarray}
 where $N$ is the number of sites,
 and $N_B$ is the number of nearest-neighbor pairs in the whole system.
 More general solutions
 for $[ \overline{e^{- \Upsilon }} ]_R$ with $h = 0$ are obtained in Refs.\cite{ON, Y}.
 By using Eq.~(\ref{eq:mu2}), we define the checking parameter $D^{(\pm J)}$
 for the $\pm J$ Ising spin glass model as 
\begin{equation}
 D^{(\pm J)} \equiv [ \overline{ \exp \{ - \Upsilon - N_B \ln [ \cosh (\beta J) ] - N \ln [ \cosh (\beta h) ] \} } ]_R \, .
\end{equation}
 The value of $D^{(\pm J)}$ is calculated
 by estimating the value of $\Upsilon$ by the Monte Carlo method.
 This checking parameter is available on all lattices, since
 $D^{(\pm J)}$ does not depend on lattice shapes although
 $D^{(\pm J)}$ depends on $N_B$ and $N$.
 $D$ is an exponential quantity for $\Upsilon$,
 and, in the Monte Carlo method,
 calculation of exponential quantities is
 generally more difficult than that of linear quantities,
 however, by examining the error bars of data,
 it is numerically seen in \S\ref{sec:4} that
 the estimation of  the quantity $D$ by the Monte Carlo method is available.

 When $D \sim 1$, the system is in global minima of the free energy
 and the system is in equilibrium or near-equilibrium.
 In other words, when $D \sim 1$, the annealing schedule is proper.
 When $D$ deviates from unity, the system is in non-equilibrium
 and/or the number of samples of realizations for exchange interaction is not enough.
 When $D \sim 0$, the system is trapped in local minima of the free energy.
 The discussion for the value of $D$ is given in \S\ref{sec:3}.
 The meaning of the value of the parameter $D$
 is different from that of the study of the energy decrease as the temperature decreases.
 The exact value of $[ \overline{e^{- \Upsilon }} ]_R$
 is analytically obtained, and
 it is checked whether the value estimated by the Monte Carlo method
 is close to the exact analytical value or not.
 This check guarantees that the system is in global minima of the free energy.

The distribution $P^{({\rm G})} ( J_{i, j})$ of $J_{i, j}$
 for the Gaussian Ising spin glass model is given by \cite{MPV}
\begin{equation}
 P^{({\rm G})} ( J_{i, j})
 = \frac{1}{\sqrt{2 \pi J^2}} \, \exp \biggl( - \frac{J^2_{i, j} }
 {2 J^2 } \biggr)  \, .
 \label{eq:PGaussianJij}
\end{equation}
 When $\beta_0 = 0$ and $\beta_\tau = \beta$, 
 by applying Eq.~(\ref{eq:Jeft1}) to the Gaussian model, we obtain
\begin{eqnarray}
 [ \overline{e^{- \Upsilon }} ]^{(G)}_R &=& \frac{1}{(2 \pi J)^{N_B / 2}} \int^{\infty}_{- \infty}
 ( \prod_{\langle i, j \rangle } J_{i, j} ) \, e^{- \frac{1}{2 J^2} \sum_{\langle i, j \rangle } J^2_{i, j} }
 \nonumber \\
 & & \times \frac{\sum_{\{ S_i \} }
 e^{\beta \sum_{\langle i , j \rangle } J_{i, j } S_i S_j  + \beta h \sum_i S_i } }{2^N} \nonumber \\
 &=& \exp \{ [ N_B (\beta J)^2 / 2 ] + N \ln [ \cosh (\beta h) ]  \} \, .
\end{eqnarray}
More general solutions
 for $[ \overline{e^{- \Upsilon }} ]_R$ with $h = 0$
 in the Gaussian model are obtained in Ref.\cite{Y}.
 We define the checking parameter $D^{(G)}$ for the Gaussian model as
\begin{equation}
 D^{(G)} \equiv [ \overline{ \exp \{ - \Upsilon - [ N_B (\beta J)^2 / 2] - N \ln [ \cosh (\beta h) ] \} } ]_R \, .
\end{equation}
 $\Upsilon$ is given in Eq.~(\ref{eq:mu}).
 The value of $ D^{(G)}$ is calculated
 by estimating the value of $\Upsilon$ by the Monte Carlo method.
 This checking parameter is available on all lattices.
 The discussion for the value of $D$ is given in \S\ref{sec:3}.

 The Hamiltonian of the glassy spin model
 by Newman and Moore is given by \cite{NM}
\begin{equation}
 {\cal H}^{(NM)} = \frac{1}{3} J \sum_{i, j, k \in \bigtriangledown} S_i S_j S_k \, ,
\end{equation}
 where $\bigtriangledown$ is the set of coordinates of downward pointing-triangles of the triangular lattice.
 This model is simple but shows a glassy behavior\cite{NM}.
 The phase transition to the ordered state occurs only at zero temperature\cite{GN}.
 $Z (\beta_0 = 0)$ and $Z (\beta_\tau = \beta)$ are respectively given by
 $Z (\beta_0 = 0) = 2^N$ and $Z (\beta_\tau = \beta) = (e^{\beta J} + e^{- \beta J})^N$
 for periodic boundary conditions.
 Thus, we define the checking parameter $D^{(NM)}$ for the glassy spin model as
\begin{equation}
 D^{(NM)} \equiv \overline{ \exp \{ - \Upsilon - N \ln [ \cosh (\beta J) ] \} } \, .
\end{equation}
 $\Upsilon$ is given in Eq.~(\ref{eq:mu}).
 The value of $ D^{(NM)}$ is calculated
 by estimating the value of $\Upsilon$ by the Monte Carlo method.
 The discussion for the value of $D$ is given in \S\ref{sec:3}.
 This model has no randomness for exchange interaction.
 Thus it is expected that
 the results of the model by Newman and Moore show that
 the reason for the behavior of $D$ is
 not because of the number of samples of realizations for exchange interaction.

\section{The Breaking of the Jarzynski Equality} \label{sec:3}

Here, we discuss that the breaking of the Jarzynski equality
 in this study is induced by the system being trapped in local minima of the free energy.
 Firstly, it is discussed that
 the system, trapped in local minima of the free energy, breaks the Jarzynski equality.
 Secondly, the relationship between the value of the checking parameter $D$
 and the system in the minima of the free energy is discussed.
 Thirdly, the relationship between the value of $D$
 and the lack of the number of samples of realizations for exchange interaction
 is discussed.

We mention a part of the derivation of the Jarzynski equality
 in order to discuss the breaking of the Jarzynski equality.
 The detail of the derivation is written in Ref.\cite{C2}.
 For convenience' sake,
 we mention the Jarzynski equality for temperature-change process.
 The Jarzynski equality for temperature-change process
 is proposed in Refs.\cite{C1, OKN}.
 In a canonical ensemble, the equilibrium probability of a state $i_t$
 given a fixed inverse temperature $\beta_t$ is given by
\begin{equation}
P ( i_t | \beta_t )  = \frac{e^{- \beta_t E (i_t )} }{Z (\beta_t ) } \, . \label{eq:Pibeta}
\end{equation}
 By using the probability $P ( i_{t - 1} \,
 \raisebox{-0.2ex}{$\longrightarrow$}\hspace{-1.1em}\raisebox{1.2ex}{{\tiny $\beta_t$}}
 \, \, \, i_t )$ of making a transition between two states,
 the detailed balance is given by
\begin{equation}
 P ( i_{t - 1} \,
 \raisebox{-0.2ex}{$\longrightarrow$}\hspace{-1.1em}\raisebox{1.2ex}{{\tiny $\beta_t$}}
 \, \, \, i_t ) P ( i_{t - 1} | \beta_t ) 
 =  P ( i_{t - 1} \,
 \raisebox{-0.2ex}{$\longleftarrow$}\hspace{-1.1em}\raisebox{1.2ex}{{\tiny $\beta_t$}}
 \, \, \, i_t ) P ( i_t | \beta_t ) \, . \label{eq:detailedb}
\end{equation}
 There is a relation:
\begin{equation}
 \frac{ P ( i_0 | \beta_0 ) P ( i_0 \,
 \raisebox{-0.2ex}{$\longrightarrow$}\hspace{-1.1em}\raisebox{1.2ex}{{\tiny $\beta_1$}}
 \, \, \, i_1) \cdots P ( i_{\tau - 1} \,
 \raisebox{-0.2ex}{$\longrightarrow$}\hspace{-1.1em}\raisebox{1.2ex}{{\tiny $\beta_\tau$}}
 \, \, \, i_\tau )}
{P ( i_\tau | \beta_\tau ) P ( i_0 \,
 \raisebox{-0.2ex}{$\longleftarrow$}\hspace{-1.1em}\raisebox{1.2ex}{{\tiny $\beta_1$}}
 \, \, \, i_1) \cdots P ( i_{\tau - 1} \,
 \raisebox{-0.2ex}{$\longleftarrow$}\hspace{-1.1em}\raisebox{1.2ex}{{\tiny $\beta_\tau$}}
 \, \, \, i_\tau )}
 = e^{\Upsilon} \frac{Z (\beta_\tau )}{Z (\beta_0)} \, . \label{eq:arelation}
\end{equation}
 By using Eqs.~(\ref{eq:mu}), (\ref{eq:Pibeta}) and (\ref{eq:detailedb}),
 one can confirm that the relation (\ref{eq:arelation}) is exact.
 Then, by calculating $\overline{e^{- \Upsilon }}$,
 the Jarzynski equality for the temperature-change process is obtained as
\begin{eqnarray}
\overline{e^{- \Upsilon }} &=&
 \sum_{i_0, i_1, \ldots, i_\tau } P ( i_0 | \beta_0 ) 
 P ( i_0 \,
 \raisebox{-0.2ex}{$\longrightarrow$}\hspace{-1.1em}\raisebox{1.2ex}{{\tiny $\beta_1$}}
 \, \, \, i_1) \cdots P ( i_{\tau - 1} \,
 \raisebox{-0.2ex}{$\longrightarrow$}\hspace{-1.1em}\raisebox{1.2ex}{{\tiny $\beta_\tau$}}
 \, \, \, i_\tau ) \, e^{- \Upsilon } \nonumber \\
 &=&
  \sum_{i_0, i_1, \ldots, i_\tau } P ( i_\tau | \beta_\tau ) 
  P ( i_0 \,
 \raisebox{-0.2ex}{$\longleftarrow$}\hspace{-1.1em}\raisebox{1.2ex}{{\tiny $\beta_1$}}
 \, \, \, i_1) \cdots P ( i_{\tau - 1} \,
 \raisebox{-0.2ex}{$\longleftarrow$}\hspace{-1.1em}\raisebox{1.2ex}{{\tiny $\beta_\tau$}}
 \, \, \, i_\tau ) \,  e^{- \Upsilon + \Upsilon} \frac{Z (\beta_\tau )}{Z (\beta_0)} \nonumber \\
 &=& \frac{Z (\beta_\tau )}{Z (\beta_0)} \, . \label{eq:brelation}
\end{eqnarray}
 In Eq.~(\ref{eq:brelation}), in order to obtain the Jarzynski equality,
 $P ( i_0 | \beta_0 )$ and $P ( i_\tau | \beta_\tau )$ are implicitly used.
 $P ( i_0 | \beta_0 )$ and $P ( i_\tau | \beta_\tau )$ are equilibrium probabilities.
 If the system is in non-equilibrium after
 a state transition sequence $i_0 \to i_1 \cdots \to i_\tau$ in a Monte Carlo simulation,
 the resulting $P ( i_\tau | \beta_\tau )$ has a deviation, and
 then the Jarzynski equality can be broken.
 The occurrence of the breaking is detected by using the checking parameter. 
 It is pointed out in Ref.\cite{J1}
 that the Jarzynski equality [Eq.~(\ref{eq:Jeft0})] does not depend on
 both of the path from $\lambda_0$ to $\lambda_\tau$ and the rate at which
 the parameters are switched along the path.
 However, these properties of the Jarzynski equality for the path and the rate
 can only be held for the system going in global minima of the free energy
 because  $P ( i_0 | \beta_0 )$ and $P ( i_\tau | \beta_\tau )$ are equilibrium probabilities
 and are implicitly used when deriving the Jarzynski equality.

 We discuss the relationship between the value of the checking parameter $D$
 and the system in the minima of the free energy.
 By using Eq.~(\ref{eq:mu}), the checking parameter is written as
\begin{equation}
 D = [ \overline{e^{- \Upsilon -\ln A}} ]_R
 = [ \overline{ e^{ - \sum^{\tau - 1}_{t = 0} ( \beta_{t + 1} - \beta_t ) E ( i_t ) - \ln A } } ]_R  \ \, , \label{eq:DA}
\end{equation}
 where $A$ is the analytical solution of $[ \overline{e^{- \Upsilon }} ]_R$.
 The part of $\beta_{t + 1} - \beta_t$ is positive,
 since an annealing process is supposed.
 From Eq.~(\ref{eq:DA}), one can see that the values of
 the energies $[E (i_0 ), E (i_1), \ldots, E (i_{\tau - 1} )]$
 are important for the equality of the Jarzynski equality.
 If the system is in local minima of the free energy,
 the energy would not decrease as the temperature decreases.
 Therefore, the energy $E^{(Loc)} (i_t )$ of the system in local minima of the free energy
 can be higher than the energy $E^{(Glo)} (i_t )$ of the system
 in global minima of the free energy.
 By using the energy $E^{(Glo)} (i_t )$,  $D \sim 1$ can be obtained.
 In other words, it is considered that,
 if  the system is in equilibrium or near-equilibrium,
 the checking parameter $D$ gives a value close to one.
 In addition, it is considered that,
 if the system is in non-equilibrium,
 the checking parameter $D$ deviates from unity.
 By using the energy  $E^{(Loc)} (i_t )$,  $D \sim 0$ can be obtained.
 In other words, it is considered that,
 if the breaking of the Jarzynski equality is induced
 by the system being trapped in local minima of the free energy,
 the checking parameter $D$ gives a value close to zero.

 A special care for the number of samples of realizations
 for exchange interactions is necessary
 when calculating models having randomness for exchange interactions.
 The disorder average on the left-hand side of Eq.~(\ref{eq:mu2})
 is dominated by disorder realizations with the most negative values
 of the pseudo work $\Upsilon$.
 If the number of samples for exchange interaction is not enough,
 then both tails of the distribution for $\Upsilon$ will be missing.
 Therefore, if the number of samples of realizations for exchange interaction is not enough,
 there is also a possibility that the checking parameter $D$ deviates from unity.

Therefore,
 from the above discussion, if $D \sim 1$,
 the system is in global minima of the free energy,
 and the system is in equilibrium or near-equilibrium.
 Thus, if $D \sim 1$, the annealing schedule is proper.
 From the above discussion,
 if $D$ deviates from unity, the system is in non-equilibrium
 and/or the number of samples of realizations for exchange interaction is not enough.
 From the above discussion,
 if $D \sim 0$, the system is trapped in local minima of the free energy.

\section{The Results of Monte Carlo Simulations} \label{sec:4}

As an example of the use of the present checking parameter,
 we performed a Monte Carlo simulation of the $\pm J$
 Ising spin glass model on the simple cubic lattice with periodic boundary conditions.
 We applied the Glauber dynamics\cite{D, Og}
 which randomly updates each single spin with probability
 $\frac{e^{- \beta \Delta E}}{1 + e^{- \beta \Delta E}}$,
 where $\Delta E$ is the energy difference by updating the corresponding spin.
 We set the value of the magnetic field $h$ to zero.
 We investigated the annealing schedule of $T = N / t^a$,
 where $a$ is an adjusting parameter,
 and, in this section, $t$ is the Monte Carlo time.
 We made temperature change per one Monte Carlo step
 according to the annealing schedule of $T = N / t^a$.
 The spins are chosen at random in the initial states.
 We estimated the checking parameter $D^{(\pm J)}$.
 The system length is $12$,
 the number of sites $N$ is $1728$, and
 the number of nearest-neighbor pairs in the whole system $N_B$ is 5184.
 We investigated $a = 1.2, 1.0$ and $0.8$ respectively.
 We set $J / k_B = 1$ for simplicity,
 and we investigated $T = 1.0, 1.1, 1.2, \ldots, 4.8$ respectively.

The result for $T'$ by using data from $T = \infty$ to $T = T'$ is obtained.
 If $T'' > T'$, by using a part of data from $T = \infty$ to $T = T'$,
 obtaining the result for $T''$ is also possible.
 We obtained the results for $T = 1.1, 1.2, \ldots, 4.8$ by using parts of data
 from $T = \infty$ to $T = 1.0$.

We investigated a case that the number of samples
 of realizations for exchange interaction is enough.
 The number that we investigated is $7000$ for each $a$.
 The number of the Monte Carlo steps of the single run for $a = 1.2$ is $498$,
 and the total number of the Monte Carlo steps for $a = 1.2$ is $3486 \times 10^3$.
 The number of the Monte Carlo steps of the single run for $a = 1.0$ is $1728$,
 and the total number of the Monte Carlo steps for $a = 1.0$ is $12096 \times 10^3$.
 The number of the Monte Carlo steps of the single run for $a = 0.8$ is $11141$,
 and the total number of the Monte Carlo steps for $a = 0.8$ is $77987 \times 10^3$.

\begin{figure}[t]
\begin{center}
\includegraphics[width=0.68\linewidth]{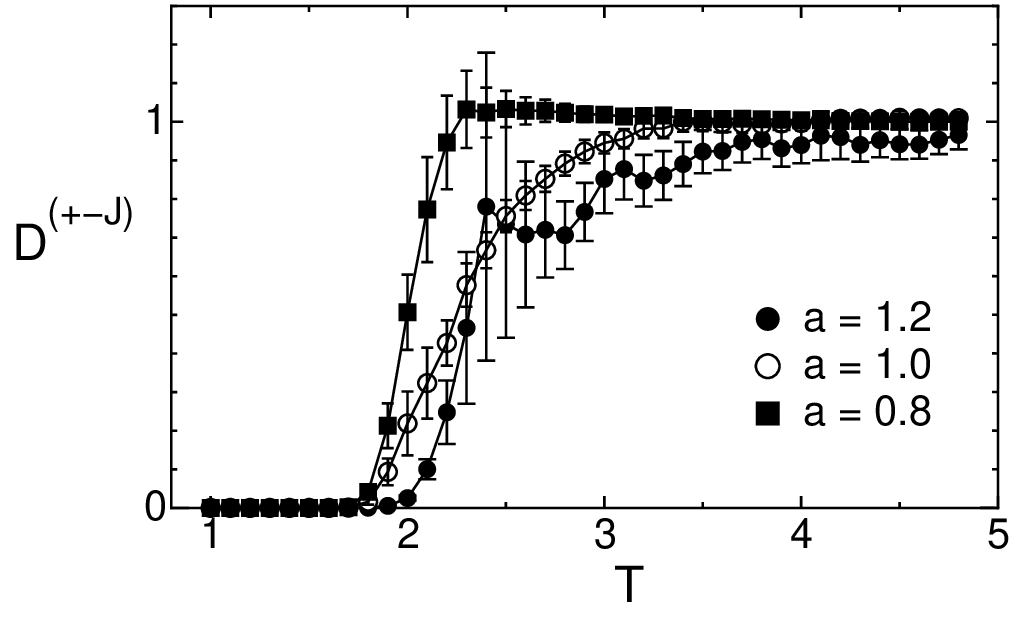}
\end{center}
\caption{
 The relation between the temperature $T$ and the checking parameter $D^{(\pm J)}$.
 The results of the $\pm J$ Ising spin glass model on the simple cubic lattice are shown.
 The system length is $12$, and the annealing schedule is $T = N / t^a$.
 The solid circle represents the result for $a = 1.2$,
 the open circle represents the result for $a = 1.0$, and
 the solid square represents the result for $a = 0.8$.
 $J / k_B = 1$ and $h = 0$ are set.
\label{fig:phase-diagram}
}
\end{figure}
 Fig.~\ref{fig:phase-diagram} shows
 the relation between the temperature $T$ and the checking parameter $D^{(\pm J)}$.
 The solid circle represents the result for $a = 1.2$,
 the open circle represents the result for $a = 1.0$, and
 the solid square represents the result for $a = 0.8$.
 As the error bars, the standard errors are shown in Fig.~\ref{fig:phase-diagram}.
 For the obtained results,
 we can see whether
 the values of $D$ are close to one or close to zero within the error bars.
 Therefore, it is numerically seen that the estimation of the quantity $D$ is available.

 The result at $a = 1.2$ shows
 that most of the values of $D^{(\pm J)}$ are not close to one
 over the whole temperature range.
 Therefore, from the discussion in \S\ref{sec:3},
 the system is considered to be in non-equilibrium over the whole temperature range.
 Also, the result for $T \le 1.8$ at $a = 1.2$ 
 shows that $D^{(\pm J)} \sim 0$.
 From the discussion in \S\ref{sec:3},
 the system is considered to be trapped in local minima of the free energy for $T \le 1.8$.

 The result for $T > 3.1$ at $a = 1.0$
 shows that $D^{(\pm J)} \sim 1$.
 Therefore, from the discussion in \S\ref{sec:3},
 the system is considered to be in equilibrium or near-equilibrium for $T > 3.1$.
 In other words,
 the annealing schedule is considered to be appropriate for $T > 3.1$.
 The result for $T \le 3.1$ at $a = 1.0$
 shows that the values of $D^{(\pm J)}$ deviate from unity.
 Therefore, from the discussion in \S\ref{sec:3},
 the system is considered to be in non-equilibrium for $T \le 3.1$.
 Also, the result for $T \le 1.7$ at $a = 1.0$
 shows that $D^{(\pm J)} \sim 0$.
 From the discussion in \S\ref{sec:3},
 the system is considered to be trapped in local minima of the free energy for $T \le 1.7$.

 The result for $T > 2.2$ at $a = 0.8$
 shows that $D^{(\pm J)} \sim 1$.
 Therefore, from the discussion in \S\ref{sec:3},
 the system is considered to be
 in equilibrium or near-equilibrium for $T > 2.2$.
 In other words,
 the annealing schedule is considered to be appropriate for $T > 2.2$.
 The result for $T \le 2.2$ at $a = 0.8$
 shows that the values of $D^{(\pm J)}$ deviate from unity.
 Therefore, from the discussion in \S\ref{sec:3},
 the system is considered to be in non-equilibrium for $T \le 2.2$.
 Also, the result for $T \le 1.7$ at $a = 0.8$
 shows that $D^{(\pm J)} \sim 0$.
 From the discussion in \S\ref{sec:3},
 the system is considered to be trapped in local minima of the free energy for $T \le 1.7$.

 In this article, since the Glauber dynamics is applied,
 discussions for dynamical features as a physical system are possible\cite{D, Og}.
 In Fig.~\ref{fig:phase-diagram}, we can see that
 the present three results converge on $T \sim 1.7$ as the temperature decreases,
 therefore it is anticipated that there is a dynamical transition point at $T \sim 1.7$.
 It is pointed out by Ogielski \cite{Og} that
 there is a dynamical transition point at $T \approx 1.8$.
 This result by Ogielski
 is seen in the non-exponential asymptotic decay of relaxation functions\cite{Og}.
 It is also pointed out by Derrida \cite{D} that
 there is a dynamical transition point at $T \sim 1.8$
 by measuring the distance between two configurations
 subjected to the same thermal noise.
 The present results may indicate the dynamical transition point.

 We performed a Monte Carlo simulation of the glassy spin model
 by Newman and Moore.
 We applied the Glauber dynamics\cite{D, Og}
 to the model with periodic boundary conditions.
 We investigated the annealing schedule of $T = N / t^a$.
 The spins are chosen at random in the initial states.
 We estimated the checking parameter $D^{(MN)}$.
 The system length is $32$, and the number of sites $N$ is $1024$.
 We investigated $a = 1.2, 1.0$ and $0.8$ respectively.
 We set $J / k_B = 1$ for simplicity,
 and we investigated $T = 0.1, 0.2, 0.3, \ldots, 3.5$ respectively.

The number of samples is $100$ for each $a$.
 The number of the Monte Carlo steps of the single run for $a = 1.2$ is $2197$,
 and the total number of the Monte Carlo steps for $a = 1.2$ is $2197 \times 10^2$.
 The number of the Monte Carlo steps of the single run for $a = 1.0$ is $10239$,
 and the total number of the Monte Carlo steps for $a = 1.0$ is $10239 \times 10^2$.
 The number of the Monte Carlo steps of the single run for $a = 0.8$ is $103008$,
 and the total number of the Monte Carlo steps for $a = 0.8$ is $103008 \times 10^2$.

\begin{figure}[t]
\begin{center}
\includegraphics[width=0.68\linewidth]{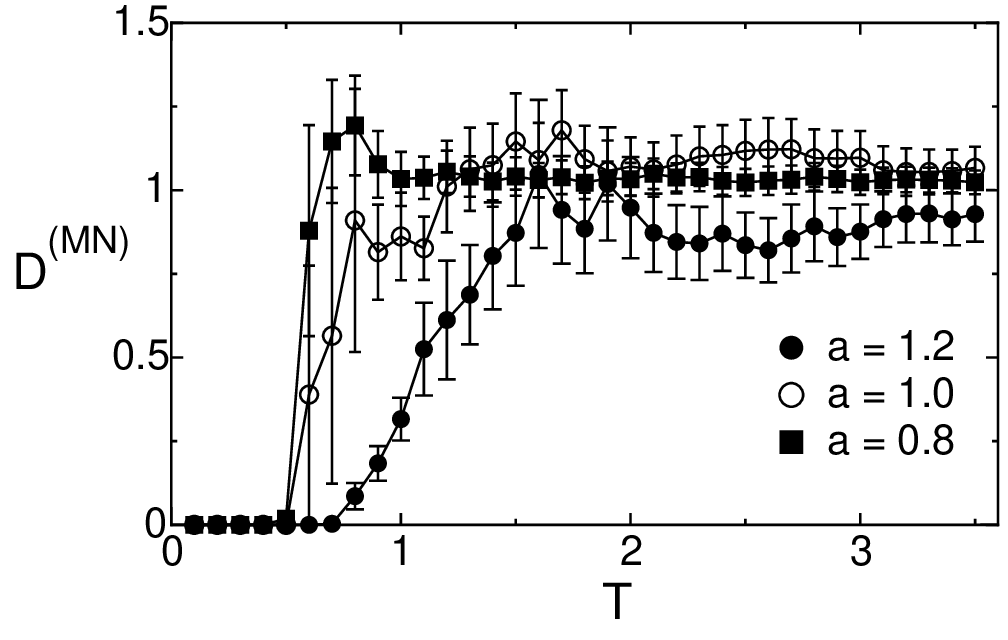}
\end{center}
\caption{
 The relation between the temperature $T$ and the checking parameter $D^{(MN)}$.
 The results of a glassy spin model proposed by Newman and Moore are shown.
 The system length is $32$, and the annealing schedule is $T = N / t^a$.
 The solid circle represents the result for $a = 1.2$,
 the open circle represents the result for $a = 1.0$, and
 the solid square represents the result for $a = 0.8$.
 $J / k_B = 1$ is set.
\label{fig:phase-diagram-2}
}
\end{figure}
 Fig.~\ref{fig:phase-diagram-2} shows
 the relation between the temperature $T$ and the checking parameter $D^{(MN)}$.
 The solid circle represents the result for $a = 1.2$,
 the open circle represents the result for $a = 1.0$, and
 the solid square represents the result for $a = 0.8$.
 As the error bars, the standard errors are shown.

 The result at $a = 1.2$ shows
 that most of the values of $D^{(MN)}$ are not close to one
 over the whole temperature range.
 Therefore, from the discussion in \S\ref{sec:3},
 the system is considered to be in non-equilibrium over the whole temperature range.
 Also, the result for $T \le 0.7$ at $a = 1.2$ shows that $D^{(MN)} \sim 0$.
 From the discussion in \S\ref{sec:3},
 the system is considered to be trapped in local minima of the free energy for $T \le 0.7$.

 The result for $T > 1.1$ at $a = 1.0$
 shows that $D^{(MN)} \sim 1$.
 Therefore, from the discussion in \S\ref{sec:3},
 the system is considered to be in equilibrium or near-equilibrium for $T > 1.1$.
 The result for $T \le 1.1$ at $a = 1.0$
 shows that the values of $D^{(MN)}$ deviate from unity.
 Therefore, from the discussion in \S\ref{sec:3},
 the system is considered to be in non-equilibrium for $T \le 1.1$.
 Also, the result for $T \le 0.5$ at $a = 1.0$
 shows that $D^{(MN)} \sim 0$.
 From the discussion in \S\ref{sec:3},
 the system is considered to be trapped in local minima of the free energy for $T \le 0.5$.

 The result for $T > 0.8$ at $a = 0.8$
 shows that $D^{(MN)} \sim 1$.
 Therefore, from the discussion in \S\ref{sec:3},
 the system is considered to be
 in equilibrium or near-equilibrium for $T > 0.8$.
 The result for $T \le 0.8$ at $a = 0.8$
 shows that the values of $D^{(MN)}$ deviate from unity.
 Therefore, from the discussion in \S\ref{sec:3},
 the system is considered to be in non-equilibrium for $T \le 0.8$.
 Also, the result for $T \le 0.5$ at $a = 0.8$
 shows that $D^{(MN)} \sim 0$.
 From the discussion in \S\ref{sec:3},
 the system is considered to be trapped in local minima of the free energy for $T \le 0.5$.

 Since the glassy spin model proposed by Newman and Moore 
 has no randomness for exchange interaction,
 this model does not suffer from the number of samples of realizations
 for exchange interaction.
 The results of this model show that
 the reason for $D$ go to zero is not because of the number of samples of realizations
 for exchange interaction.

\section{Concluding Remarks} \label{sec:5}

We proposed a checking parameter utilizing the breaking of the Jarzynski equality 
 in the simulated annealing method using the Monte Carlo method.
 This parameter is based on the Jarzynski equality.
 By using this parameter, to detect that
 the system is in global minima of the free energy
 under gradual temperature reduction is possible.
 Thus, by using this parameter, one is able
 to investigate the efficiency of annealing schedules.
 We applied this parameter to the $\pm J$ Ising spin glass model.
 The application to the Gaussian Ising spin glass model was also mentioned.
 We discussed that the breaking of the Jarzynski equality in this study is induced
 by the system being trapped in local minima of the free energy.
 As an example, we performed a Monte Carlo simulation of the $\pm J$
 Ising spin glass model and showed  the efficiency of the use of this parameter.

 In addition, in order to confirm the behavior of the present parameter,
 a glassy spin model \cite{NM} by Newman and Moore was also investigated.
 Although a special care for the number of samples of realizations for exchange interactions
 is necessary when calculating models having randomness for exchange interactions,
 the glassy spin model has no randomness for exchange interaction, and
 the results of the model by Newman and Moore showed that
 the reason for $D$ go to zero is not because of the number of samples of realizations
 for exchange interaction.

 In this article, we applied the Glauber dynamics\cite{D, Og}.
 The Glauber dynamics is suited to the study
 of the dynamical features of physical systems\cite{D, Og}.
 If one uses the present checking parameter as a powerful optimization method,
 applying more powerful Monte Carlo methods\cite{HN, N},
 instead of the Glauber dynamics, would be necessary.
 The investigation for applying more powerful Monte Carlo methods
 is a future task.

 There is a problem whether a reasonable annealing schedule
 to find the ground state exists or not.
 The present result has shown that there is a reasonable annealing schedule
 for a certain temperature range.
 The reasonable schedule to find the ground state is not known.
 If an annealing schedule was analytically derived,
 the present checking parameter would judge whether the schedule is effective or not.

The checking parameter for
 a quantum annealing\cite{FGGS, RCC, FGSSD, KN} can also be obtained.
 The Hamiltonian is given by \cite{OKN}
${\cal H}^{(QA)} = - \lambda \sum_{\langle i, j \rangle} J_{i, j} \sigma^z_i \sigma^z_j - (1 - \lambda ) \sum_i \sigma^x_i$, where $\sigma^k_i$ is the $k$ component of Pauli matrix at site $i$,
 $\lambda$ is switched from $0$ to $1$,
 and this switching of $\lambda$ corresponds to the annealing process.
 The distribution of $J_{i, j}$ is given in Eq.~(\ref{eq:PpmJJij}).
  $[ \overline{e^{- \beta W }} ]_R$ is analytically obtained \cite{OKN} as $[ \overline{e^{- \beta W }} ]_R = [\cosh (\beta J)]^{N_B} / [\cosh (\beta )]^N$. Thus
 we define the checking parameter $D^{(QA)}$
 for the quantum annealing as
 $D^{(QA)} \equiv [ \overline{ e^{ - \beta W - \ln \{ [\cosh (\beta J)]^{N_B} / [\cosh (\beta )]^N   \} } } ]_R $.
 $W$ is given in Eq.~(\ref{eq:W}).
 The value of $ D^{(QA)}$ is calculated
 by estimating the value of $W$
 for the partition function decomposed by the Suzuki-Trotter decomposition\cite{S}
 or the high-temperature series expansion\cite{H} by the Monte Carlo method.
 This checking parameter is available on all lattices.
 When $D \sim 1$, the system is in global minima of the free energy
 and is in equilibrium or near-equilibrium.
 In other words, when $D \sim 1$, the annealing schedule is proper.
 When $D$ deviates from unity, the system is in non-equilibrium
 and/or the number of samples of realizations for exchange interaction is not enough.
 When $D \sim 0$, the system is trapped in local minima of the free energy.
 The discussion for the value of $D$ is given in \S\ref{sec:3}.

The checking parameters for investigating other models can also be obtained.
 In order to obtain the checking parameter used in a model,
 it can be necessary that the equilibrium information
 in the Jarzynski equality is analytically solved.

\end{document}